\documentclass[twocolumn,superscriptaddress,aps]{revtex4-1}
\usepackage{graphicx}
\usepackage{amsmath}
\usepackage{amsthm}
\usepackage{amssymb}
\usepackage{latexsym}
\usepackage{array}
\usepackage{hyperref}
\usepackage{float}
\usepackage{amsfonts}
\usepackage{mathrsfs}
\usepackage{verbatim}
\usepackage{bbold}

\usepackage{makecell}

\usepackage{color}

\newcommand{\bra}[1]{\left<#1\right|}
\newcommand{\ket}[1]{\left|#1\right>}
\newcommand{\abs}[1]{\left|#1\right|}

\newtheorem{theorem}{Theorem}
\newtheorem{definition}{Definition}

\begin{document}

\title{Quantum secure learning with classical samples}

\author{Wooyeong~Song}
\thanks{The first two authors contributed equally to this work}
\affiliation{Department of Physics, Hanyang University, Seoul 04763, Korea}

\author{Youngrong~Lim}
\thanks{The first two authors contributed equally to this work}
\affiliation{School of Computational Sciences, Korea Institute for Advanced Study, Seoul 02455, Korea}

\author{Hyukjoon~Kwon}
\affiliation{QOLS, Blackett Laboratory, Imperial College London, London SW7 2AZ, United Kingdom}

\author{Gerardo Adesso}
\affiliation{School of Mathematical Sciences and Centre for the Mathematics and Theoretical Physics of Quantum Non-Equilibrium Systems, University of Nottingham, University Park, Nottingham NG7 2RD, United Kingdom}

\author{Marcin~Wie\'{s}niak}
\affiliation{Institute of Theoretical Physics and Astrophysics, Faculty of Mathematics, Physics and Informatics, University of Gda\'{n}sk, 80-308 Gda\'{n}sk, Poland}
\affiliation{International Centre for Theory of Quantum Technologies, University of Gda\'{n}sk, 80-308 Gda\'{n}sk, Poland}

\author{Marcin~Paw\l{}owski}
\affiliation{International Centre for Theory of Quantum Technologies, University of Gda\'{n}sk, 80-308 Gda\'{n}sk, Poland}

\author{Jaewan~Kim}
\affiliation{School of Computational Sciences, Korea Institute for Advanced Study, Seoul 02455, Korea}

\author{Jeongho~Bang}\email{jbang@etri.re.kr}
\affiliation{Electronics and Telecommunications Research Institute, Daejeon 34129, Korea}

\received{\today}

\begin{abstract}
Studies addressing the question ``Can a learner complete the learning securely?'' have recently been spurred from the standpoints of fundamental theory and potential applications. In the relevant context of this question, we present a classical-quantum hybrid sampling protocol and define a security condition that allows only legitimate learners to prepare a finite set of samples that guarantees the success of the learning; the security condition excludes intruders. We do this by combining our security concept with the bound of the so-called probably approximately correct (PAC) learning. We show that while the lower bound on the learning samples guarantees PAC learning, an upper bound can be derived to rule out adversarial learners. Such a secure learning condition is appealing, because it is defined only by the size of samples required for the successful learning and is independent of the algorithm employed. Notably, the security stems from the fundamental quantum no-broadcasting principle. No such condition can thus occur in any classical regime, where learning samples can be copied. Owing to the hybrid architecture, our scheme also offers a practical advantage for implementation in noisy intermediate-scale quantum devices.
\end{abstract}

\maketitle

%---------------------------------------------------------------------------------------------------------------------------------------------------------------------------------------------------------
\section{Introduction}
%---------------------------------------------------------------------------------------------------------------------------------------------------------------------------------------------------------

The hybridization of machine learning and quantum theory has been intensively studied, especially to explore the possibility of exploiting quantum learning speedups. Very recently, the incorporation of useful quantum-algorithm-kernel (e.g., quantum linear solvers~\cite{Harrow2009}) into data processing tasks in machine learning has yielded encouraging results~\cite{Rebentrost2014,Lloyd2014,Schuld2016,Kerenidis2016}. Within a span of a few years, such approaches have become increasingly important in quantum computation, leading to the advent of quantum machine learning~\cite{Biamonte2017,Schuld2015}.

In parallel, the issue of security has been of considerable interest to the machine learning community. The term ``secure learning'' is usually used to indicate that the learning is allowed only for the legitimate learner, who wants to rule out adversarial learners. The main objective of these adversaries is to acquire ability to become equals of the legitimate learner or to render the learning of the legitimate learner counterproductive. In this context, one of the open issues is how to define a secure learning condition for detecting and preventing these adversaries. While this problem has been widely studied in classical learning~\cite{Barreno2010,Nelson2012}, only a few quantum mechanical studies have been conducted so far~\cite{Bang2015,Sheng2017,Liu2018}.

We indicate that the legitimate learning mates can communicate a (classically) encrypted dataset after generating a secret key via a well-established quantum-key-distribution (QKD) scheme. In that case, it would be impractical for the adversarial learner(s) to extract critical learning information once the QKD is completed. However, the adversarial learner(s) may want to spoil the learning by disrupting the communication. Such a purpose can be achieved simply by disrupting the encrypted data after the key is distributed. This is actually one of the distinctive aspects of the learning security~\cite{Barreno2010}. Thus, the learning security can neither be fully achieved nor defined by the QKD alone.

Having the above in mind, we in this paper construct a secure learning condition with favorable quantum properties. To this end, we first design a protocol for secure sampling that runs between two legitimate learning parties. We cast a classical-quantum hybrid oracle that allows large-size classical inputs with a small-scale quantum system~\cite{Harrow2020}. As the main result, we derive a secure learning condition such that only the original legitimate learner is guaranteed success for learning; we designate the condition as the {\em secure probably-approximately-correct (PAC) learning} condition. The beauty of this condition is that the security is derived only from the size of learning samples the legitimate learner requires and it stems from the quantum no-broadcasting principle~\cite{Barnum1996,Dang2007}. Therefore, such condition cannot be defined in any classical regime. Our paper also leads to an intriguing classical-quantum interplay, namely, in which the (large) input data remain classical while the useful quantum properties are explored for a small quantum system~\cite{Lee2019,Song2019}. Such architecture helps avoid the use of a largely superposed sample and is well suited to noisy intermediate-scale quantum (NISQ) technologies~\cite{Preskill2018}.

%---------------------------------------------------------------------------------------------------------------------------------------------------------------------------------------------------------
\section{Problem}

Given a (Boolean) function $c \in C$ that maps the input $\mathbf{x}=x_0 x_1 \cdots x_{n-1}$ to a binary value $c(\mathbf{x}) \in \{0, 1\}$, learning is defined as the process of identifying a hypothesis $h \in {\cal H}$ close to $c$. The binary number $x_j \in \{0, 1\}$ ($j=0,1,\ldots,n-1$) can be considered as the ``feature'' and the size of the hypothesis set $\abs{\cal H}$, called ``model complexity,'' is assumed to be finite. Such a problem covers a wide variety of learning tasks. In particular, this binary setting of the problem can, in principle, be extended to a more general situation such as multi-class tasks~\cite{Rifkin2004}. For this reason, the binary classification framework has generally been used in computational learning theory~\cite{Kearns1994,Langley1995}. 

In such a problem, the learner, say Alice ($\mathscr{A}$), should first sample a set $T$ of input-target pairs, where $T=\{ (\mathbf{x}, c(\mathbf{x})) \}$. To accomplish this sampling, $\mathscr{A}$ employs a black box, called the oracle. The oracle is responsible for accessing critical information, namely, $c(\mathbf{x})$ for a given $\mathbf{x}$. Here, we assume that the oracle is owned by $\mathscr{A}$'s distant partner, say Bob ($\mathscr{B}$). Such an assumption, namely, of the two learning parties being located far apart, is commonly invoked in secure learning~\cite{Barreno2010}. The issue is then how $\mathscr{A}$ can sample a clean dataset $T$ with $\mathscr{B}$ in a manner that is secure against any malicious attack; in other words, how can $\mathscr{A}$ learn $c$ securely?

%---------------------------------------------------------------------------------------------------------------------------------------------------------------------------------------------------------
\section{Secure sampling protocol}
%---------------------------------------------------------------------------------------------------------------------------------------------------------------------------------------------------------

\begin{figure}[t]
\centering
\includegraphics[width=0.46\textwidth]{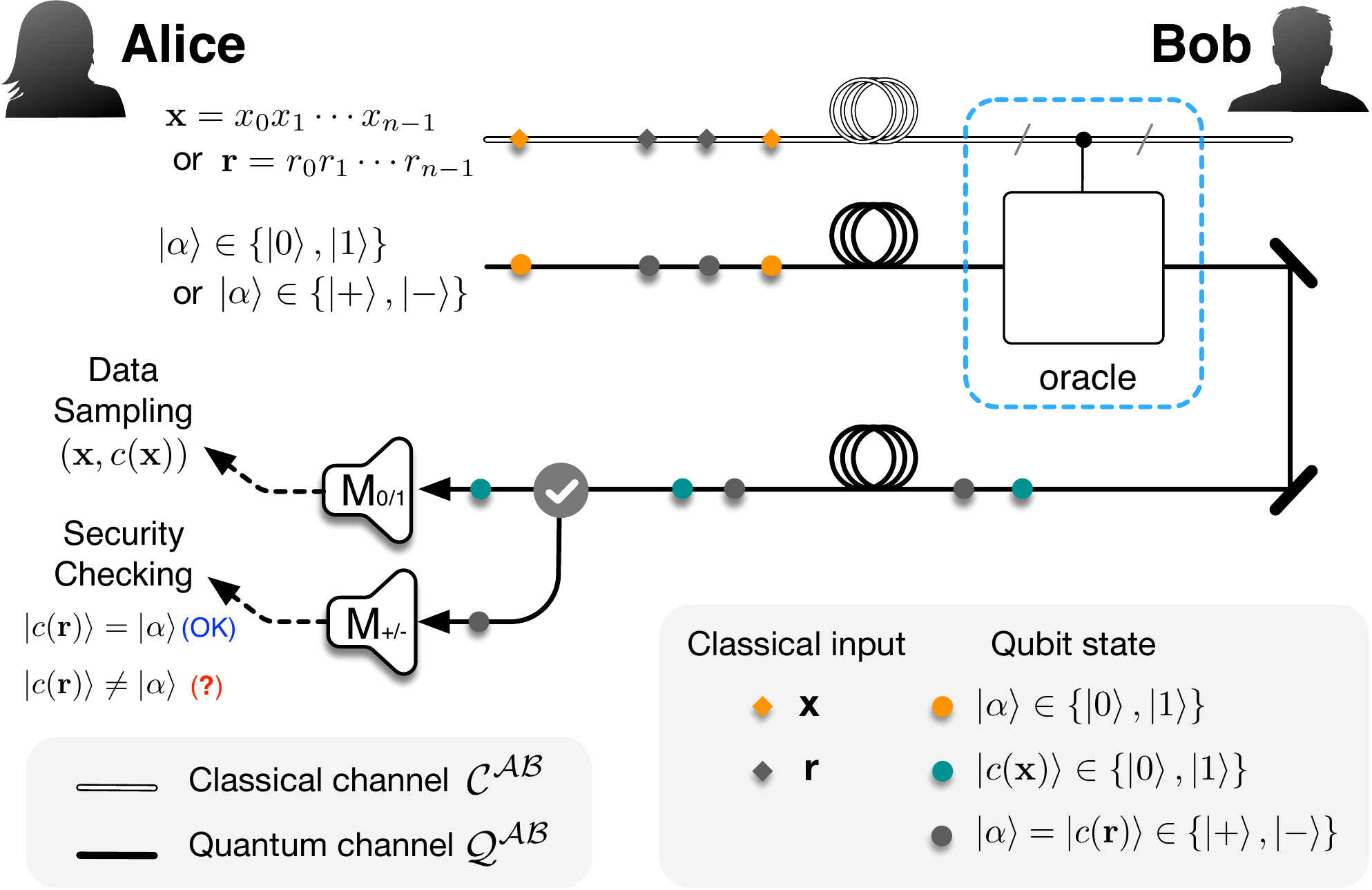}
\caption{\label{fig:protocol} Schematic of our sampling protocol. Alice ($\mathscr{A}$) has facilities for the preparation of inputs, $(\mathbf{x}, \ket{\alpha} \in \{\ket{0}, \ket{1}\})$ or $(\mathbf{r}, \ket{\alpha} \in \{\ket{+}, \ket{-}\})$. $\mathscr{A}$ can also perform a single-qubit measurement to identify the returning qubit. Bob ($\mathscr{B}$) owns the oracle. Here, we consider a classical-quantum hybrid architecture (blue dashed and solid boxes) with a classical input ($\mathbf{x}$ or $\mathbf{r}$) and an ancillary qubit state ($\ket{\alpha}$). The oracle does not reveal its structure. $\mathscr{A}$ and $\mathscr{B}$ communicate via classical and quantum channels, denoted by ${\cal C}^\mathscr{AB}$ and ${\cal Q}^\mathscr{AB}$, respectively.}
\end{figure}

We introduce a classical-quantum hybrid oracle $O(c)$, which consists of input and output channels for $n$-bit classical data $\mathbf{x}$ and for a single qubit, denoted by ${\cal C}^\mathscr{AB}$ and ${\cal Q}^\mathscr{AB}$, respectively. This oracle $O(c)$ implements $(\mathbf{x}, \ket{\alpha}) \to (\mathbf{x}, \ket{c(\mathbf{x}) \oplus \alpha})$ for $\alpha \in \{0,1\}$ and $(\mathbf{r}, \ket{\alpha}) \to (\mathbf{r}, \ket{\alpha})$ for $\alpha \in \{+,-\}$, where $\ket{c(\mathbf{x})}$ is the oracle-answer for a given $\mathbf{x}$. Here, $\mathbf{r}$ is a {\em random} input which is casted for the purpose of testing the existence of any malicious intruder who disturbs the communication. Thus, $\mathbf{r}$ is chosen such that $(\mathbf{r}, y) \notin T$ (for any $y \in \{0, 1\}$). The construction of such an operation is fairly common, e.g., in QKD or quantum secure direct communication schemes~\cite{Scarani2009,Hu2016}. Note that it is not permissible to extract any information by looking into $O(c)$. A useful hybrid oracle architecture is presented in Appendix~\ref{appendix:A}. 

We now present the secure sampling protocol, which proceeds as follows. First, $\mathscr{A}$ prepares the state $\ket{\alpha}$ as an eigenstate of $\hat{\sigma}_z$ or $\hat{\sigma}_x$ (i.e., $\ket{\alpha} \in \{ \ket{0}, \ket{1}, \ket{\pm}=\frac{1}{\sqrt{2}}\left( \ket{0} \pm \ket{1} \right) \}$) at random. The prepared state $\ket{\alpha}$ is transferred to $\mathscr{B}$ through ${\cal Q^\mathscr{AB}}$. If $\ket{\alpha} = \ket{0}$ or $\ket{1}$, $\mathscr{A}$ sends the input $\mathbf{x}$ through ${\cal C^\mathscr{AB}}$ together with $\ket{\alpha}$, and if $\ket{\alpha} = \ket{\pm}$, $\mathscr{A}$ draws a random input $\mathbf{r}$. Subsequently, ($\mathbf{x}$, $\ket{\alpha} \in \{\ket{0}, \ket{1}\}$) or ($\mathbf{r}$, $\ket{\alpha} \in \{\ket{+}, \ket{-}\}$) are passed through the oracle $O(c)$, and the output states $\ket{c(\mathbf{x})}$ or $\ket{\pm}$ of the qubit are returned to $\mathscr{A}$. For $\ket{\alpha} \in \{ \ket{0}, \ket{1} \}$, $\mathscr{A}$ obtains a sample pair ($\mathbf{x}$, $c(\mathbf{x})$) by performing $\hat{\sigma}_z$ measurement, and for $\ket{\alpha} \in \{ \ket{+}, \ket{-} \}$, $\mathscr{A}$ should receive $\ket{\alpha}=\ket{\pm}$ from $\mathscr{B}$. Therefore, by checking the returned state $\ket{\pm}$ with the $\hat{\sigma}_x$ measurement, $\mathscr{A}$ can sense any adversarial learner, often referred to as Eve ($\mathscr{E}$), who alters the qubits moving $\mathscr{A}\to\mathscr{B}$ or $\mathscr{B}\to\mathscr{A}$ (see Fig.~\ref{fig:protocol}). Note that $(\mathbf{r}, y) \notin T$ for any $y \in \{0,1\}$ obtained by $\hat{\sigma}_z$ measurement, and it cannot be a valid sample.

%---------------------------------------------------------------------------------------------------------------------------------------------------------------------------------------------------------
\section{No-broadcasting of learning samples}
%---------------------------------------------------------------------------------------------------------------------------------------------------------------------------------------------------------

\begin{figure}[t]
\centering
\includegraphics[width=0.46\textwidth]{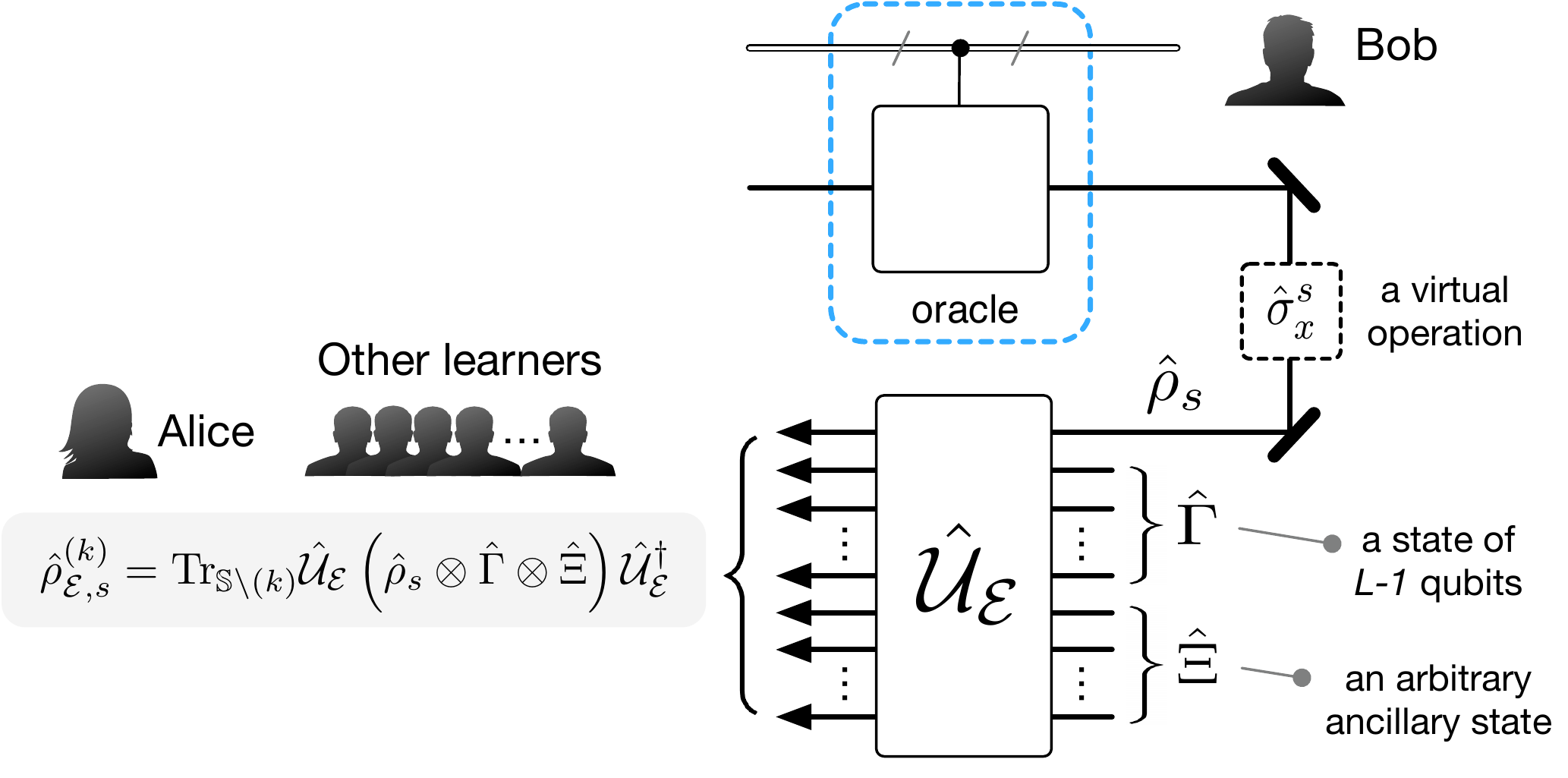}
\caption{\label{fig:e_strategy} General attack by adversarial learners. Here, we consider $L-1$ adversarial learners who can freely access ${\cal C}^{\mathscr{AB}}$ and ${\cal Q}^{\mathscr{AB}}$. Each adversarial learner has his or her own (in principle, infinite size) ancillary system and is assumed to be an expert in quantum theory. We further assume that the adversarial learners can team up to process an optimal strategy ${\cal E}$ for their own or for the group's benefit.}
\end{figure}

With the protocol described above, we present our first result:
\begin{theorem}
In our protocol, for any given $c \in C$, $\mathscr{B}$ cannot distribute the full set of learning samples, namely, $T=\{(\mathbf{x}, c(\mathbf{x}))\}$, to $\mathscr{A}$ and other (external) learners. Therefore, the condition
\begin{eqnarray}
T = T^{(k)}~(\forall k \in [1,L]),
\label{eq:broadcasting}
\end{eqnarray}
where $T^{(k)}$ is the set of samples that the $k$ learner (i.e., $\mathscr{A}$ or $\mathscr{E}$) finally gets for strategy $\mathcal{E}$, cannot be satisfied.
\label{thr:1}
\end{theorem}
For proving this theorem, we let $\hat{\rho}_0=\ket{c(\mathbf{x})}\bra{c(\mathbf{x})}$ and $\hat{\rho}_1 = \ket{\alpha}\bra{\alpha}$, each of which is defined in terms of a state of the ideal oracle output in a trial for a given input ($\mathbf{x}$ or $\mathbf{r}$). Here, $\hat{\rho}_0 \in \{\ket{0}\bra{0}, \ket{1}\bra{1}\}$ and $\hat{\rho}_1 \in \{\ket{+}\bra{+}, \ket{-}\bra{-}\}$. Suppose $\mathscr{B}$ adopts a strategy $\mathcal{E}$ to distribute the samples in $T$ among learners (including $\mathscr{A}$), with $L \ge 2$. In general, $\mathcal{E}$ can be represented as a completely positive and trace-preserving map with an overall unitary $\hat{\mathcal{U}}_{\mathcal{E}}$ and an arbitrary ancilla state $\hat{\Xi}$ (see Fig.~\ref{fig:e_strategy}). The distributed states can be written such that
\begin{eqnarray}
\hat{\rho}_{\mathcal{E},s}^{(k)} = \text{Tr}_{\mathbb{S} \backslash (k)} \hat{\mathcal{U}}_{\mathcal{E}} \left( \hat{\rho}_s \otimes \hat{\Gamma} \otimes \hat{\Xi} \right) \hat{\mathcal{U}}_{\mathcal{E}}^\dagger, 
\end{eqnarray}
where $\text{Tr}_{\mathbb{S} \backslash (k)}$ denotes the partial trace with respect to all systems $\mathbb{S}$ except the one labeled with the $k$th learner, and $\hat{\Gamma}$ represents a state of $L-1$ qubits, each of which is distributed to the corresponding learner, except $\mathscr{A}$ (here, $k=1$ denotes $\mathscr{A}$). Then, it is true that $\mathscr{B}$ cannot {\em broadcast} the states $\hat{\rho}_s$ ($s=0,1$) to a ($k$-indexed) learner. This is confirmed by the principle that the states $\hat{\rho}_0$ and $\hat{\rho}_1$ are not distinguishable~\cite{Barnum1996,Chefles1998,Barnum2007}. Therefore, a sample pair ($\mathbf{x}$, $c(\mathbf{x})$) cannot be shared for a given $\mathbf{x}$. Thus, the full set of $T$ cannot be distributed in the complete form and Theorem~\ref{thr:1} holds.

%---------------------------------------------------------------------------------------------------------------------------------------------------------------------------------------------------------
\section{Secure Probably-Approximately-Correct Learning}
%---------------------------------------------------------------------------------------------------------------------------------------------------------------------------------------------------------

Suppose that $\mathscr{A}$ is the only legitimate learner, and the other $L-1$ learners are malicious intruders. Without loss of generality, we let $k \in \{\mathscr{A}, \mathscr{E}\}$ with $L=2$, or equivalently, by assuming that all $L-1$ intruders team up together as one $\mathscr{E}$. In this setting, we can assume that $\mathcal{E}$ is a general attack strategy adopted by $\mathscr{E}$. Then, Theorem~\ref{thr:1} describes the following situation: if $\mathscr{E}$ disturbs the protocol, the samples prepared by $\mathscr{A}$ (and also $\mathscr{E}$) must be {\em noisy}; specifically, a portion $\eta^\mathscr{A}$ (and $\eta^\mathscr{E}$) of the contaminated samples, for example, $(\mathbf{x}, c(\mathbf{x}) \oplus 1)$, would be included in $\mathscr{A}$'s (and $\mathscr{E}$'s) samples. Note that $\mathscr{A}$ and $\mathscr{E}$ cannot identify these contaminations. Here, $\eta^{(k)} \le \frac{1}{2}$ ($k \in \{\mathscr{A},\mathscr{E}\}$) and is determined by $\mathscr{E}$'s strategy ${\cal E}$. It can be written as (for $T \gg 1$)
\begin{eqnarray}
\eta^{(k)} = 1 - \frac{\abs{T_S^{(k)}}}{\abs{T^{(k)}}} \ge 1 - \min_{s} F(\hat{\rho}_s, \hat{\rho}_{\mathcal{E},s}^{(k)}), 
\label{eq:eta_k}
\end{eqnarray}
where $T_S^{(k)}$ denotes the set of uncontaminated samples in $T^{(k)}$; thus, $T_S^{(k)} \subseteq T^{(k)}$ and $T_S^{(k)} \subseteq T$. $F(\hat{\rho}, \hat{\sigma})$ is the fidelity between the states $\hat{\rho}$ and $\hat{\sigma}$~\cite{Jozsa1994}. Here, the inequality in the rightmost side is introduced because $\mathscr{E}$ would make the contaminated samples even in cases where $\hat{\rho}_s$ are correctly cloned~\footnote{This is because $\mathscr{E}$ is unaware of which are the communication (or testing) rounds.}. The equality is always saturated for $\mathscr{A}$. We then assume that our protocol forbids any strategy ${\cal E}$ that allows the condition
\begin{eqnarray}
\left( \eta_c > \eta^{\mathscr{A}} \right) \land \left( \eta_c > \eta^{\mathscr{E}} \right)
\label{eq:no_bc-th1}
\end{eqnarray}
with a critical factor $\eta_c$. This assumption is true when $\eta_c$ is chosen such that $\eta_c = 1 - F_\text{opt}$ where $F_\text{opt}$ is the optimal fidelity achievable by a ($1 \to 2$) $\hat{\rho}_s$ cloner~\footnote{For example, $\eta_c=\frac{1}{6}$ in the qubit case~\cite{Dang2007}.}. Then, Eq.~(\ref{eq:no_bc-th1}) can be rewritten by using Eq.~(\ref{eq:eta_k}) as 
\begin{eqnarray}
\left( F_\text{opt} < F(\hat{\rho}_s, \hat{\rho}_{\mathcal{E},s}^{\mathscr{A}}) \right) \land \left( F_\text{opt} < F(\hat{\rho}_s, \hat{\rho}_{\mathcal{E},s}^{\mathscr{E}}) \right),
\end{eqnarray}
which immediately contradicts the quantum no-cloning principle~\cite{Dang2007}. We note that if Alice could acquire information about Eve's attack scenario (if any), it might be possible to consider a more useful $\eta_c$ setting. If $\eta_c = 0$, Eq.~(\ref{eq:no_bc-th1}) becomes equivalent to the condition Eq.~(\ref{eq:broadcasting}) and we encounter Theorem~\ref{thr:1}.

We now discuss secure learning in the framework of the so-called PAC learning~\cite{Valiant1984,Langley1995}. In a PAC learning, the concept class $C$ is said to be $(\epsilon, \delta)$-PAC learnable [we call the learner a $(\epsilon, \delta)$-PAC learner] if an $\epsilon$-approximated correct solution (i.e., hypothesis) $h \in H$ can be found with a probability $1-\delta$; in other words, $C$ is said to be $(\epsilon, \delta)$-PAC learnable if $P[E(h, c) \le \epsilon] \ge 1-\delta$ is satisfied for any $c \in C$, where $E(h, c)$ is an error function that indicates how $h$ and $c$ differ~\cite{Langley1995}. Such a theorem of PAC learning indicates that if a learner is allowed to use a certain size, say $M_b(\epsilon, \delta)$, of contaminated samples with $\eta$, he or she is guaranteed to be a $(\epsilon, \delta)$-PAC learner. In this case, $\eta$ is defined as the percentage of contaminated samples in the entire set of samples [refer to Eq.~(\ref{eq:eta_k})]. Usually, $M_b(\epsilon, \delta)$ is referred to as ``sample complexity''~\cite{Angluin1994,Kearns1994}. Here, $M_b(\epsilon, \delta)$ is divided into two categories depending on whether the samples are ideal (i.e., $\eta=0$) or noisy (i.e., $\eta \in (0, \frac{1}{2}]$) (For more details, see Appendix~\ref{appendix:B}, Refs.~\cite{Valiant1984,Langley1995}, and the informative summary in Chap.~$5$ of Ref.~\cite{Ciliberto2018}). The latter, namely, the noisy PAC learning model, provides a useful framework and is suitable for our paper because contaminations, either from $\mathscr{E}$ or from imperfection intrinsic to the channels, can be included in the expression for $\eta^{(k)}$. 

It is noteworthy that the (full) quantum model of the PAC learning, namely, quantum PAC learning, was also developed by using a quantum oracle that allows the (large) superposition of the inputs $\mathbf{x}$~\cite{Ciliberto2018}. However, no study has been conducted on secure learning in a classical or a quantum PAC learning framework.

We now present our second result:
\begin{theorem}
For any given $c \in C$, let $M_b^{\mathscr{A}}(\epsilon, \delta)$ and $M_b^{\mathscr{E}}(\epsilon, \delta)$ denote the ``optimal'' sample complexities of $\mathscr{A}$ and $\mathscr{E}$, respectively~\footnote{By 'optimal' we mean that $M_b^{(k)}(\epsilon, \delta)$ ($k \in \{\mathscr{A}, \mathscr{E}\}$) is the minimum size of learning samples necessary and sufficient such that there exists a PAC $k$-learner requiring at most $M_b^{(k)}(\epsilon, \delta)$ samples for any given $\eta^{(k)}$~\cite{Hanneke2016}.}. Then, during the running of our protocol, if $\mathscr{A}$ becomes a $(\epsilon, \delta)$-PAC learner by identifying the samples smaller than $M_b^{\mathscr{E}}(\epsilon, \delta)$, $\mathscr{E}$ cannot become a $(\epsilon, \delta)$-PAC learner for the same $\epsilon$ and $\delta$.
\label{thr:2}
\end{theorem}
The proof of this theorem is as follows. First, consider the case $\eta^{\mathscr{A}} \ge \eta^{\mathscr{E}}$, which will lead to $M_b^{\mathscr{A}}(\epsilon, \delta) \ge M_b^{\mathscr{E}}(\epsilon, \delta)$. In this case, it is impossible for $\mathscr{A}$ to be a $(\epsilon, \delta)$-PAC learner with $M$ samples smaller than $M_b^{\mathscr{E}}(\epsilon, \delta)$. Second, in the case of $\eta^{\mathscr{A}} < \eta^{\mathscr{E}}$, if $\mathscr{A}$ completes the learning with $M$ samples and becomes a $(\epsilon, \delta)$-PAC learner satisfying $M_b^{\mathscr{E}}(\epsilon, \delta) > M \ge M_b^{\mathscr{A}}(\epsilon, \delta)$, then $\mathscr{E}$ cannot simultaneously be a $(\epsilon, \delta)$-PAC learner because the protocol will be terminated before $\mathscr{E}$ obtains a sufficient number of samples (i.e., larger than $M_b^{\mathscr{E}}(\epsilon, \delta)$) to be a $(\epsilon, \delta)$-PAC learner. This proves Theorem~\ref{thr:2}.

On the basis of the above analysis, we present a definition for a secure learner:
\begin{definition}
For any $c \in C$, suppose $\mathscr{A}$ identifies $h$ with $M$ samples, with
\begin{eqnarray}
M_c(\epsilon, \delta) \ge M \ge M_b(\epsilon, \delta).
\label{eq:SPAC_condi}
\end{eqnarray}
Here, $M_b(\epsilon, \delta)$ and $M_c(\epsilon, \delta)$ are defined as $M_b^{(k)}(\epsilon, \delta)$ when $\eta^{(k)} \to 0$ and $\eta^{(k)} \to \eta_c$, respectively, where $k$ is either $\mathscr{A}$ or $\mathscr{E}$. Then, we call $\mathscr{A}$ a quantum secure $(\epsilon, \delta)$-PAC learner.
\label{def:1}
\end{definition}
In this definition, the lower bound of the sample size [i.e., $M \ge M_b(\epsilon, \delta)$] is necessary for $\mathscr{A}$ to be a $(\epsilon, \delta)$-PAC learner. The upper bound [i.e., $M_c(\epsilon, \delta) \ge M$] is adopted for security, and it follows from Theorem~\ref{thr:2} and Eq.~(\ref{eq:no_bc-th1}).

For wide applicability of Theorem~\ref{thr:1}, \ref{thr:2} and Definition~\ref{def:1}, we apply two additional rules: ({\bf R.1}) When the number of trials for ($\mathbf{r}$, $\ket{\alpha}$) reaches ${M_b(\epsilon, \delta) - \Gamma}$, then $\mathscr{A}$ tests whether $\frac{M_{c(\mathbf{r})\neq \alpha}}{M_b(\epsilon, \delta) - \Gamma}$ is larger than ${\eta_c - \Delta}$, where $M_{c(\mathbf{r})\neq \alpha}$ is the number of inconsistent results [i.e., $c(\mathbf{r}) \neq \alpha$] in $\mathscr{A}$'s $\hat{\sigma}_x$ measurement. If $\frac{M_{c(\mathbf{r})\neq \alpha}}{M_b(\epsilon, \delta)  - \Gamma} \ge \eta_c - \Delta$, $\mathscr{A}$ suspends the process by confirming that the state change, namely, $\ket{\pm} \to \ket{\mp}$, occurs by $\mathscr{E}$; otherwise, $\mathscr{A}$ continues the process. Here, we approximate 
\begin{eqnarray}
\eta^{\mathscr{A}} \simeq \frac{M_{c(\mathbf{r})\neq \alpha}}{M_b(\epsilon, \delta)}
\end{eqnarray}
by assuming $M_{c(\mathbf{x}) \to c(\mathbf{x}) \oplus 1} = M_{c(\mathbf{r})\neq \alpha}$, where $M_{c(\mathbf{x}) \to c(\mathbf{x}) \oplus 1}$ denotes the number of contaminated pairs in $\mathscr{A}$'s sample set after a certain number of trials. This assumption is reasonable because $\mathscr{A}$ generates ($\mathbf{r}$, $\ket{\alpha} \in \{\ket{+}, \ket{-}\}$) or $(\mathbf{x}, \ket{\alpha} \in \{\ket{0}, \ket{1}\})$ with probability $\frac{1}{2}$, which cannot be discriminated by $\mathscr{E}$. ({\bf R.2}) If the learning is not completed until the number of trials for ($\mathbf{x}$, $\ket{\alpha}$) reaches $M_c(\epsilon, \delta)$, $\mathscr{A}$ quits the process. It is to be noted that the factors $\Gamma$ and $\Delta$ in ({\bf R.1}) are introduced to limit the quality of $\mathscr{E}$'s learning.

We can now analyze the possible situations. First, let us consider the case (i) $\eta^{\mathscr{A}} \ge \eta^{\mathscr{E}}$. Then, the following two subcases can be considered:
\begin{eqnarray}
\text{(i-a) $\eta^{\mathscr{A}} \ge \eta_c - \Delta \ge \eta^{\mathscr{E}}$ and (i-b) $\eta^{\mathscr{A}} \ge \eta^{\mathscr{E}} \ge \eta_c - \Delta$}. \nonumber
\end{eqnarray}
However, cases (i-a) and (i-b) do not actually happen because ({\bf R.1}) will halt the process when $\eta^{\mathscr{A}} \ge \eta_c - \Delta$; hence $\mathscr{E}$ is not allowed to become a $(\epsilon, \delta)$-PAC learner. Second, for the case (ii) $\eta^{\mathscr{A}} < \eta^{\mathscr{E}}$, we can also consider the following two subcases:
\begin{eqnarray}
\text{(ii-a) $\eta^{\mathscr{E}} > \eta_c - \Delta \ge \eta^{\mathscr{A}}$ and (ii-b) $\eta^{\mathscr{E}} > \eta^{\mathscr{A}} \ge \eta_c - \Delta$}. \nonumber
\end{eqnarray}
In case (ii-a), if $\mathscr{A}$ can learn $h \simeq c$ (for any given $\epsilon$ and $\delta$) with $M$ samples, with $M$ satisfying Eq.~(\ref{eq:SPAC_condi}), $\mathscr{A}$ becomes a secure $(\epsilon, \delta)$-PAC learner according to Definition~\ref{def:1}, while $\mathscr{E}$ cannot. However, {\em at least in theory}, it is not impossible for $\mathscr{E}$ to obtain the samples with a size identical to $\mathscr{A}$'s after the completion of $\mathscr{A}$'s learning. Nevertheless, $\mathscr{E}$ cannot be a ($\epsilon$, $\delta$)-PAC learner at the same level as $\mathscr{A}$ since $\eta^{\mathscr{E}}$ cannot be smaller than $\eta^{\mathscr{A}} + \Delta$. The condition $\eta^{\mathscr{A}} \ge \eta_c - \Delta$ in (ii-b) will also halt the protocol because of rule ({\bf R.1}). Thus, our results (i.e., Theorem~\ref{thr:1} and \ref{thr:2} and Definition~\ref{def:1}) can be practically applied to the protocol against any $\mathscr{E}$. Further, by using $\Gamma$ and $\Delta$, we can set the minimum gap between the level of $\mathscr{A}$'s and $\mathscr{E}$'s PAC learning in the worst case, and it would prevent $\mathscr{E}$ from becoming a slightly weaker PAC learner than $\mathscr{A}$. The subcases $\eta_c - \Delta \ge \eta^{\mathscr{A}} \ge \eta^{\mathscr{E}}$ and $\eta_c - \Delta \ge \eta^{\mathscr{E}} > \eta^{\mathscr{A}}$ are not expected to occur since they contradict Eq.~(\ref{eq:no_bc-th1}).

%---------------------------------------------------------------------------------------------------------------------------------------------------------------------------------------------------------
\section{Multi-class classification}
%---------------------------------------------------------------------------------------------------------------------------------------------------------------------------------------------------------

We also consider the multi-class problem by assuming that the input $\mathbf{x}$ belongs to $2^m$ different classes ($m \ge 2$). Here, we briefly sketch two strategies: 

(i) First, the multi-class classification problem is commonly solved by decomposing it into several binary problems. For instance, the ``one-vs-all (OVA)'' reduction is often used~\cite{Rifkin2004}, where the problem is decomposed into $2^m$ decisions of $h_i$ ($i \in \{0,1,\ldots, 2^m -1\}$) that separates the learning data of the $i$th class from the other ones. Then, datum $\mathbf{x}$ is classified with $\arg \max_i h_i (\mathbf{x})$. Here, the condition for secure PAC learning in Eq.~(\ref{eq:SPAC_condi}) can be applied to each decision of $h_i$. However, a long learning time is required because the condition in Eq.~(\ref{eq:SPAC_condi}) should be satisfied for every $2^m$ decisions. 

(ii) In another way, we can consider a single-machine approach, where the oracle can answer for all $2^m$ labels, that is, $\mathbf{y} \in \{0,1\}^m$, by allowing $m$ qubits conditioned by the same $\mathbf{x}$-input channels. In such generalization, our theorems and the condition in Eq.~(\ref{eq:SPAC_condi}) can also be applied consistently for the states of an arbitrary number of qubits. However, in this case, the region that satisfies the secure PAC learning, i.e., $\abs{M_c(\epsilon, \delta) - M_b(\epsilon, \delta)}$, narrows. In other words, the security condition becomes more stringent. For detailed analysis, see Appendix~\ref{appendix:C}.

%---------------------------------------------------------------------------------------------------------------------------------------------------------------------------------------------------------
\section{Remarks}
%---------------------------------------------------------------------------------------------------------------------------------------------------------------------------------------------------------

We have presented a concept of secure learning that safeguards against any malicious manipulation of learning samples. In contrast to other studies on secure learning, we constructed an analytic framework based on a computational model of learning theory, called PAC learning. This allowed us to establish the link between sample complexity and the condition for learning security. Our approach is appealing because the security condition is defined solely by the sample size; in particular, it is independent of $\mathscr{A}$'s (or $\mathscr{E}$'s) learning algorithms.

Our derivations of Theorem~\ref{thr:1} and \ref{thr:2} were based on the quantum principle of no-broadcasting of states, and using these theorems, we introduced the concept of secure PAC learning. Such a security condition cannot exist in the classical regime where $\mathscr{E}$ can create as many copies of the learning samples as he or she wishes.

It is noteworthy that our protocol was designed based on a classical-quantum hybridization, where the input data remain classical but only a single-qubit system is employed. Such a hybridization differs considerably from those of other hybrid models. This architecture renders our protocol suitable for NISQ implementation, without the requirement of an excessively large superposition of samples and/or without accessing a novel quantum gadget, called quantum random-access memory~\cite{Giovannetti08-1,Giovannetti08-2}.

We finally point out that determining a more practical form of $M_c(\epsilon, \delta)$ in Eq.~(\ref{eq:SPAC_condi}) continues to be an open problem, and it will be considered in a follow-up study. Notably, it is related to the determination of the optimal sample complexity, which has been a long-standing interest in computational learning theory, especially in the case where the samples are noisy. We believe that our paper will contribute to expanding the frontiers for quantum secure machine learning.

%---------------------------------------------------------------------------------------------------------------------------------------------------------------------------------------------------------
\section*{Acknowledgements}
W.S., Y.L. and J.B. are grateful to Nana Liu for valuable discussions. This work was partly supported by National Research Foundation of Korea (NRF) grants (2019R1A2C2005504, NRF-2019M3E4A1079666, and 2020M3E4A1079939), funded by the MSIP (Ministry of Science, ICT and Future Planning) and Institute of Information \& communications Technology Planning \& Evaluation (IITP) grant funded by the Korea government (MSIT) (No.~2020-0-00890, ``Development of trusted node core and interfaces for the interoperability among QKD protocols''). W.S., Y.L., and J.B. acknowledge the research project on developing quantum machine learning and quantum algorithm (No.~2019-100) by the ETRI affiliated research institute. Y.L. acknowledges National Research Foundation of Korea a grant funded by the Ministry of Science and ICT (NRF-2020M3E4A1077861) and KIAS Individual Grant (CG073301) at Korea Institute for Advanced Study. H. K. is supported by the KIST Open Research Program. M.P. and M.W. acknowledge the ICTQT IRAP project of FNP, financed by structural funds of EU. G.A. acknowledges financial support from the European Research Council (ERC) under the Starting Grant GQCOP (Grant No.~637352) and the Foundational Questions Institute (FQXi) under the Intelligence in the Physical World Programme (Grant No.~FQXiRFP-IPW-1907). M.W. was supported by NCN grants 2015/19/B/ST2/01999 and 2017/26/E/ST2/01008. M.P. was supported under FNP grant First Team/2016-1/5. J.K. was supported in part by KIAS Advanced Research Program (No.~CG014604). J.B. was also supported by a KIAS Individual Grant (No.~CG061003).

%-------------------------------
\appendix
%-------------------------------

%---------------------------------------------------------------------------------------------------------------------------------------------------
\section{Useful Classical-Quantum Hybrid Oracle Architecture}\label{appendix:A}
%---------------------------------------------------------------------------------------------------------------------------------------------------

Here, we present an example of a classical-quantum hybrid oracle, which can be applied to our study of secure learning. This oracle allows the classical inputs $\mathbf{x}$ and a single qubit $\ket{\alpha}$. It performs the mapping
\begin{eqnarray}
\left( \mathbf{x}, \ket{\alpha} \right) \to \left( \mathbf{x} , \ket{c(\mathbf{x}) \oplus \alpha} \right)~\text{for}~\alpha \in \{0,1\},
\end{eqnarray}
and
\begin{eqnarray}
\left( \mathbf{r}, \ket{\alpha} \right) \to \left( \mathbf{r} , \ket{\alpha} \right)~\text{for}~\alpha \in \{+,-\},
\end{eqnarray}
where $\mathbf{r}$ is a random datum that is to be used for performing a security check. Note that $\mathbf{x}$ remains unaltered during and after the sampling process. 

\begin{figure}[t]
\includegraphics[width=0.46\textwidth]{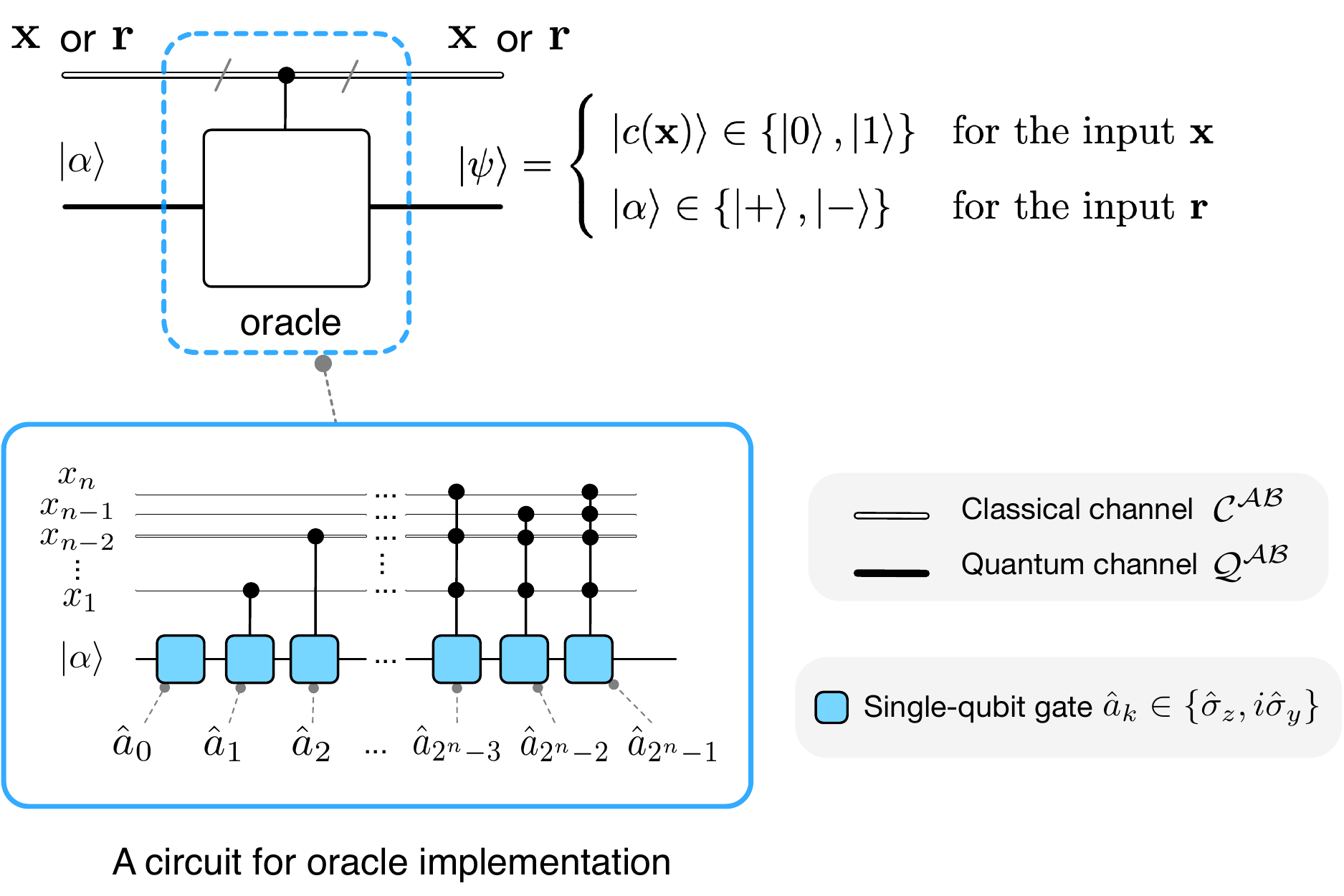}
\caption{\label{fig:oracle} {\bf Schematic of a hybrid oracle.} The oracle consists of two different input and output channels: classical input data $\mathbf{x} = x_1 x_2 \cdots x_n$ ($x_j \in \{0,1\}$ $\forall j=1,\ldots,n$) and a single qubit to produce the oracle output states. This oracle applies $2^n$ unitary gates $\hat{a}_k \in \{ \hat{\sigma}_z, i\hat{\sigma}_y \}$ ($k=0,1,\ldots,2^n-1$) conditioned on the values of the classical bits $x_j$ in $\mathbf{x}$ to the qubit channel. In a purely classical case, these gates are either identity or logical-NOT gates.}
\end{figure}

This hybrid oracle can be implemented by a circuit having a specific architecture, such as that shown in Fig.~\ref{fig:oracle}. This circuit contains $2^n$ gates acting on the ancilla qubit: the single-qubit gate $\hat{a}_0$ and $2^{n}-1$ gates $\hat{a}_k$ ($k=1,2,\ldots,2^{n}-1$) conditioned on the classical-bit values $x_1, x_2, \ldots, x_n$ in $\mathbf{x}$. The gates $\hat{a}_k$ are given by
\begin{eqnarray}
\hat{a}_k \in \left\{ \hat{\sigma}_z, i \hat{\sigma}_y \right\},~\text{for all}~k=0,1, \ldots, 2^n -1,
\label{eq:gates}
\end{eqnarray}
where $\hat{\sigma}_x$, $\hat{\sigma}_y$, and $\hat{\sigma}_z$ are the Pauli operators. This architecture of the oracle is inspired by the general expression of a Boolean function~\cite{Gupta06}, which is given by
\begin{eqnarray}
h^\star(\mathbf{x}) &=& a_0 \oplus a_1 x_1 \oplus a_2 x_2 \oplus a_3 x_1 x_2 \nonumber \\
    && \oplus \cdots \oplus a_{2^n -1} x_1 x_2 \ldots x_n,
\label{eq:univ_c}
\end{eqnarray}
where $a_k \in \{0, 1\}$ ($k=0,1,\ldots,2^n-1$) are known as the Reed--Muller coefficients. Here, each coefficient has a corresponding gate operation $\hat{a}_k$. More specifically, $a_k =0$ implies that $\hat{a}_k$ leaves the bit signal unchanged (identity), while $a_k=1$ indicates that $\hat{a}_k$ flips the bit signal (logical-not)~\cite{Toffoli80}. The oracle is thus characterized by a fixed set of gates $\hat{a}_k$ for a given $c$. Information on the gates $\hat{a}_k$ and how they run is not provided, and it should be learned. Such an oracle architecture indeed differs from other hybrid schemes. It has been argued that such hybridization can offer the advantage of being NISQ implementable and of achieving speedups~\cite{Lee2019,Song2019}.

%---------------------------------------------------------------------------------------------------------------------------------------------------
\section{Probably-Approximately-Correct (PAC) Learning Model}\label{appendix:B}
%---------------------------------------------------------------------------------------------------------------------------------------------------

In the PAC learning model~\cite{Valiant1984}, a learner samples a finite set of training data $\{ (\mathbf{x}_i, c(\mathbf{x}_i)) \}$ ($i = 1,2,\ldots,M$) by accessing an oracle. Here, $\mathbf{x}_i$ is typically assumed to be drawn uniformly. For any $c \in {\cal C}$, a learning algorithm is a ($\epsilon$, $\delta$)-PAC learner (under uniform distribution) if it can obtain an $\epsilon$-approximated correct $h \in {\cal H}$ with probability $1-\delta$. More specifically, a learning algorithm is a ($\epsilon$, $\delta$)-PAC learner if it satisfies the condition
\begin{eqnarray}
\text{Prob}(E(h, c) \le \epsilon) \ge 1-\delta,
\label{eq:pac_e}
\end{eqnarray}
where $E(h, c)$ denotes the error, for example, the distance between $h$ and $c$. If the obtained $h$ agrees with
\begin{eqnarray}
M \ge \frac{1}{\epsilon}\ln{\frac{{\abs{\cal H}}}{\delta}}
\label{eq:sample_C}
\end{eqnarray}
of samples constructed from the oracle, then Eq.~(\ref{eq:pac_e}) holds. Here, $\abs{\cal H}$ denotes the cardinality of ${\cal H}$, often-called model complexity. In the standard context, Eq.~(\ref{eq:sample_C}) is known as the ``sample complexity''~\cite{Valiant1984,Langley1995}. In other words, it yields the minimum number of training samples required to successfully learn $h \in {\cal H}$, satisfying Eq.~(\ref{eq:pac_e}). Such a sample complexity derived from previous {\em classical} studies can be directly used in our scenario. In our classical-quantum hybrid query scheme, the same sample complexity exists since $\mathbf{x}_i$ and $c(\mathbf{x}_i)$ identified by the measurement performed by Alice are classical. The beauty of this theorem is that the condition for being a PAC learner depends only on the number of samples, not on any specific learning algorithm.

In the case where the oracle outputs are contaminated, the sample complexity in Eq.~(\ref{eq:sample_C}) is modified as follows: First, we draw a sequence of training data,
\begin{eqnarray}
 \{ (\mathbf{x}_1, m_1), (\mathbf{x}_2, m_2), \ldots, (\mathbf{x}_M, m_M) \},
\end{eqnarray}
where $m_i \in \{ c(\mathbf{x}_i), c(\mathbf{x}_i) \oplus 1 \}$ denotes the outcome of the measurement performed by Alice. Subsequently, if sampling is performed with 
\begin{eqnarray}
M \ge \frac{2 \xi}{\epsilon^2}} \ln{\left(\frac{2\abs{\cal H}}{\delta}\right),
\label{eq:sc_q}
\end{eqnarray}
we can verify that Eq.~(\ref{eq:pac_e}) holds for the algorithm that obtains $h \in {\cal H}$. It has been proven that the additional factor $\xi$ is given by~\cite{Angluin1994}
\begin{eqnarray}
\xi = \frac{1}{\left( 1 - 2\eta \right)^{2}}.
\label{eq:factor_Aq1}
\end{eqnarray}
Such a {\em noisy} PAC learning model provides a useful framework for our study of secure learning. It is noteworthy that in our scenario, the contamination of the output because of an attack by Eve and that resulting from imperfections intrinsic to the oracle can be incorporated together into the factor $\eta$.

%---------------------------------------------------------------------------------------------------------------------------------------------------
\section{Extension to multi-class classification}\label{appendix:C}
%---------------------------------------------------------------------------------------------------------------------------------------------------

Each training datum can be considered to belong to one of $2^m$ different classes ($m \ge 2$), and the goal is to learn a hypothesis that, given a (new) data point, can correctly decide the class to which the data point belongs. This problem is called the multi-class classification problem.

\subsection{One-vs-All (OVA) reduction} %---------------------------------------

\begin{figure}[t]
\includegraphics[width=0.46\textwidth]{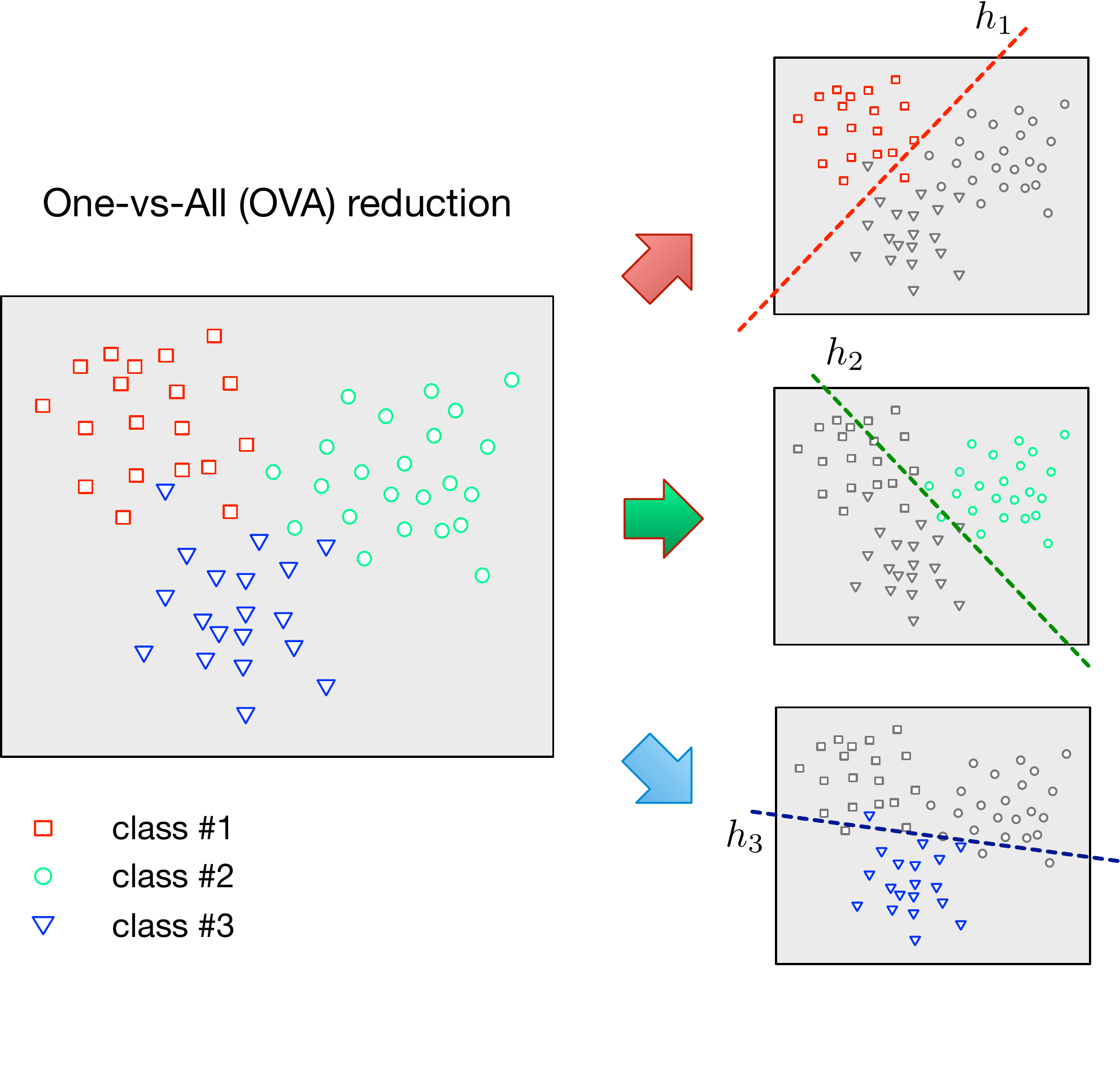}
\caption{\label{fig:OVA} Schematic of OVA reduction for three classes.}
\end{figure}

The conventional approach used to solve the multi-class classification problem is to decompose the problem into several binary classification problems. The most simple, but powerful, method is the so-called OVA reduction~\cite{Rifkin2004}, where each binary classifier (e.g., RLSC, SVM) is trained to distinguish the examples in a single class from those in all remaining classes. More specifically, in such strategy, the problem is decomposed to $2^m$ decisions of $h_i$, ($i \in \{0,1,\ldots,2^m -1\}$) that separates the training data of the $i$th class from those of the other classes (see Fig.~\ref{fig:OVA}), and (new) data are classified using
\begin{eqnarray}
h(\mathbf{x}) = \arg \max_i h_i (\mathbf{x}),
\end{eqnarray}
where $h_i(\mathbf{x})$ is a hypothesis identified in each trial and $h(\mathbf{x})$ is a decision for the classification of the input $\mathbf{x}$. Here, $h_i(\mathbf{x})$ is interpreted as the probability of a given input being included in the $i$th class, which is very suitable for our PAC learning framework. To achieve OVA reduction, we can apply the condition for secure PAC learning (Eq.~(7) of our main paper), as it is, to each trial performed for identifying $h_i(\mathbf{x})$. However, in this case, the learning time is increased as we should prepare the dataset to train $2^m$ classifiers and the secure PAC learning condition should be satisfied for every $2^m$ trials.

\subsection{A strategy of single-machine approach} %---------------------------------------

\begin{figure}[t]
\includegraphics[width=0.46\textwidth]{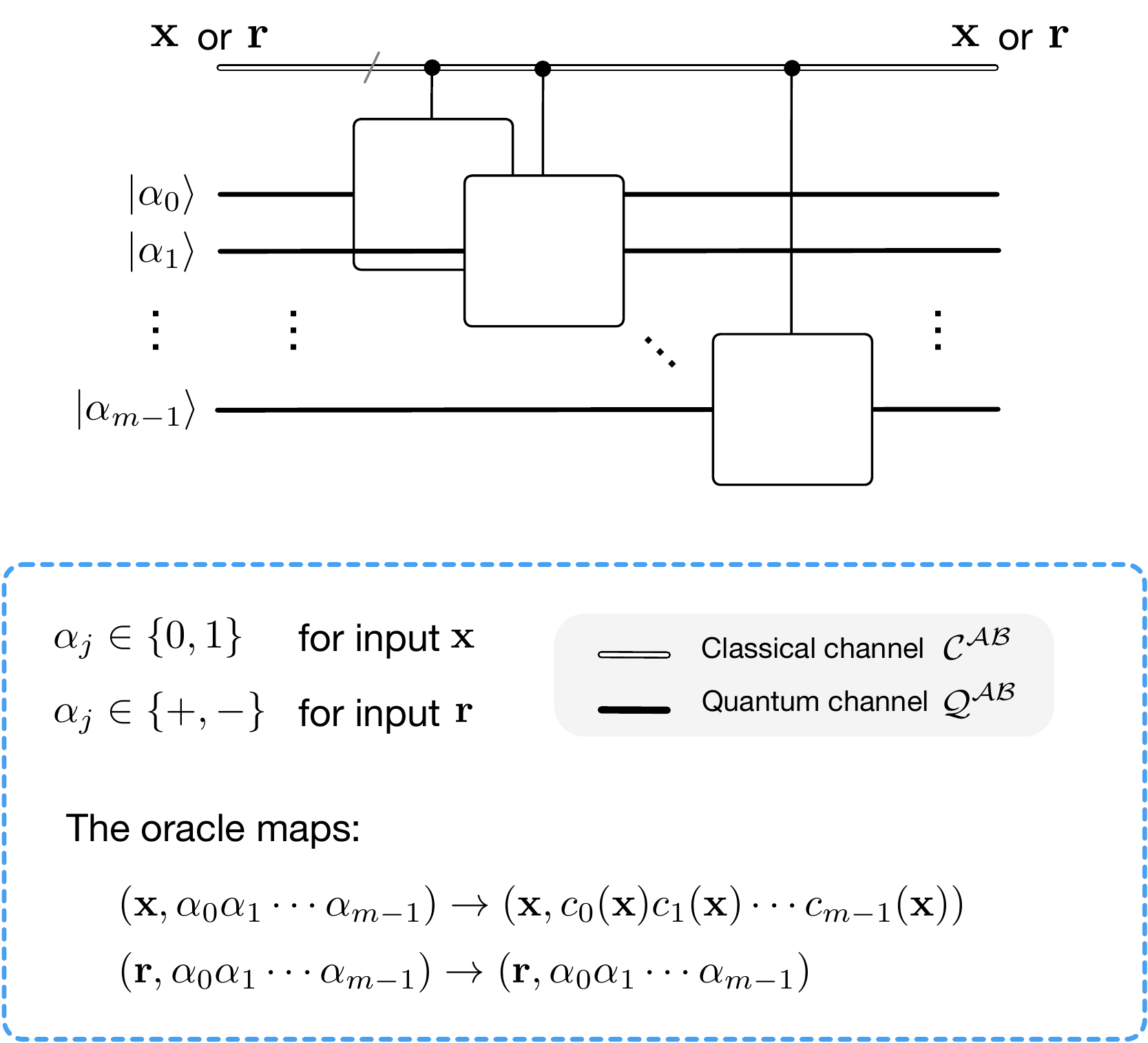}
\caption{\label{fig:sm_approach} Schematic of the oracle for a N-class output.}
\end{figure}

Another useful approach is to solve a single optimization problem that trains many binary classifiers {\em simultaneously}; this approach is akin to the so-called ``single machine approach''~\cite{Rifkin2004}. To apply this approach, we should consider an oracle that, given an input $\mathbf{x} \in \{0,1\}^n$, outputs the corresponding label $\mathbf{y} \in \{0, 1\}^m$ for all $2^m$ classes, for example, by employing an arbitrary function $h: \{ 0, 1 \}^n \to \{ 0, 1\}^m$. This is possible by allowing $m$ qubits conditioned by the same $\mathbf{x}$-input channels (see Fig.~\ref{fig:sm_approach}). More specifically, in this generalization, the oracle performs the following mapping
\begin{eqnarray}
(\mathbf{x}, \alpha_0 \alpha_1 \cdots \alpha_{m-1}) \to (\mathbf{x}, c_0(\mathbf{x}) c_1(\mathbf{x}) \cdots c_{m-1}(\mathbf{x}) )
\end{eqnarray}
for the learning (i.e., for $\alpha_0 \alpha_1 \cdots \alpha_{m-1} \in \{0,1\}^m$) and the mapping
\begin{eqnarray}
(\mathbf{r}, \alpha_0 \alpha_1 \cdots \alpha_{m-1}) \to (\mathbf{r}, \alpha_0 \alpha_1 \cdots \alpha_{m-1})
\end{eqnarray}
for the security check (i.e., for $\alpha_0 \alpha_1 \cdots \alpha_{m-1} \in \{+,-\}^m$). The learner (Alice, here) can identify the oracle's output by measuring each returning qubit and construct the training samples for the learning. In this strategy, our theorems and the secure PAC learning condition can be applied to the states of an arbitrary number of qubits. Note that in our analysis, the states $\hat{\rho}_s$ and $\hat{\rho}_{\cal E}^{(k)}$ comprise an arbitrary number of qubits. The rules [{\bf R.1}] and [{\bf R.2}] derived for practical use of our protocol are applicable to each qubit measurement outcome. However, in this case, $M_b(\epsilon, \delta)$ is expected to increase as a higher model complexity, $\abs{\cal H}$, would be imposed for large $m$. Furthermore, $M_c(\epsilon, \delta)$ decreases since $\eta_c$ increases for large $m$; specifically, we have~\cite{Chen2007}
\begin{eqnarray}
\eta_c = 1 - \max{F(\hat{\rho}_0(\mathbf{x})^{\otimes m}, \hat{\rho}_{\cal E}^{\otimes m})} = \frac{1}{(2m + 4)}.
\end{eqnarray}
Consequently, the region $\abs{M_c(\epsilon, \delta) - M_b(\epsilon, \delta)}$ that satisfies the secure PAC learning narrows as $m$ increases; in other words, the security condition becomes more stringent. Therefore, there exists a {\em trade-off} between the two aforementioned approaches. Note that $\abs{M_c(\epsilon, \delta) - M_b(\epsilon, \delta)} \ge 0$ is always satisfied along with the no-broadcasting theorem with the condition $\left( \eta_c \ge \eta^{\mathscr{A}} \right) \land \left( \eta_c \ge \eta^{\mathscr{E}} \right)$ in Eq.~(4) of our main paper.

\end{document}